\documentclass[10pt]{article}
\usepackage{infocomp}
\usepackage{times}
\usepackage{amsmath}
\usepackage{amssymb}
\usepackage[T1]{fontenc}
\usepackage[english]{babel}
\usepackage{graphicx}
\usepackage{hyperref}
\usepackage{subfigure}
\usepackage{enumerate}
\usepackage{caption}
\hypersetup{colorlinks,urlcolor=blue}
\usepackage{float}
\usepackage{rotating}
\usepackage{amsfonts}
\usepackage{footnote}
\usepackage{tablefootnote}
\usepackage{longtable}
\usepackage{array}
\usepackage{multirow}
\usepackage{graphicx}
\usepackage{epstopdf}
\usepackage[utf8]{inputenc}
\usepackage{booktabs}
\usepackage{bbding}
\usepackage{tikz}
\usetikzlibrary{shapes,arrows}
\usepackage[export]{adjustbox}
\usepackage{lineno}
\usepackage{textcomp}
\newcommand{\specialcell}[2][c]{%
  \begin{tabular}[#1]{@{}c@{}}#2\end{tabular}}

\address{Department of Computer Architecture, Universitat Polit\`ecnica de Catalunya (BarcelonaTech), Barcelona, Spain.\\
    Telecommunications and Control Engineering Department, School of Engineering, University of Sa\~o Paulo, Brazil.\\
        $^1$\url{ayyoub@ac.upc.edu},~\url{ayyoub@ieee.org}\\
        $^2$\url{jra@lcs.poli.usp.br},~\url{amazonas@ac.upc.edu}\\
        $^3$\url{german@ac.upc.edu}\\
        $^4$\url{pareta@ac.upc.edu}}

\title{Service Level Agreements for Communication Networks: A Survey}

\author{
        Ayyoub Akbari-Moghanjoughi$^1$ \\
        Jos\'e Roberto de Almeida Amazonas$^2$ \\
        Germ\'an Santos-Boada$^3$\\
        Josep Sol\'e-Pareta$^4$
}

\abstract{Information and Communication Technology (ICT) is being provided to the variety of end-users demands, thereby providing a better and improved management of services is crucial. Therefore, Service Level Agreements (SLAs) are essential and play a key role to manage the provided services among the network entities. This survey identifies the state of the art covering concepts, approaches and open problems of the SLAs establishment, deployment and management. This paper is organised in a way that the reader can access a variety of proposed SLA methods and models addressed and provides an overview of the SLA actors and elements. It also describes SLAs' characteristics and objectives. SLAs' existing methodologies are explained and categorised followed by the Service Quality Categories (SQD) and Quality-Based Service Descriptions (QSD). SLA modelling and architectures are discussed, and open research problems and future research directions are introduced. The establishment of a reliable, safe and QoE-aware computer networking needs a group of services that goes beyond pure networking services. Therefore, within the paper this broader set of services are taken into consideration and for each Service Level Objective (SLO) the related services domains will be indicated. The purpose of this survey is to identify existing research gaps in utilising SLA elements to develop a generic methodology, considering all quality parameters beyond the Quality of Service (QoS) and what must or can be taken into account to define, establish and deploy an SLA. This study is still an active research on how to specify and develop an SLA to achieve the win-win agreements among all actors.}


\receivedate{May 19th, 2019}

\acceptdate{June 1st, 2019}

\begin{document}

\maketitle

\section{Introduction}\label{sec1}
Over the years, the question of how to specify, provide and measure service quality for network end-users has been of utmost interest for service providers, their clients as well as network infrastructure providers~\cite{Richters1998}\cite{Gozdecki2003}\cite{Shetty2010}\cite{ITUQoS2008}. Furthermore, during the last decade the liberalization and deregulation process started in the telecommunication's environment. Increasing competition, favoured conjointly by client's performance needs, imposes huge pressures on service providers and network. Moreover, for many years, since having encountered particularly cost decreases, these days, in order to differentiate their product from the other competitors, providers try to improve quality of service (QoS).

Hence, duty of each and every entity that participate in service provision and their relations need to be explained. The domain is to elucidate liabilities of every supplier and to assure quality of service needed by client. The Service Level Agreement (SLA) is a beneficial tool in formalising the interrelationships among all entities, that is the consequence of a negotiation among all participating actors with the target of achieving a common comprehension concerning delivery of services, their priorities,  quality, responsibilities, and other relevant parameters \cite{itusla}\cite{Trygar2005}. To measure the agreed SLA performance among entities, plenty of monitoring tools and protocols have been developed by well-known companies. As an instance, Cisco's Service Level Assurance Protocol (Cisco's SLA Protocol) is a protocol for performance evaluation which is deployed extensively. This is utilised for service level parameters measurement such as network delay variation, latency, as well as frame and packet loss ratio. Furthermore, this protocol characterises the Cisco SLA Protocol Measurement-Type UDP-Measurement to enable the interoperability of service providers.\cite{ciscosla}. 

\begin{figure}[!htbp]
\centering
\includegraphics[draft=false,width=3in]{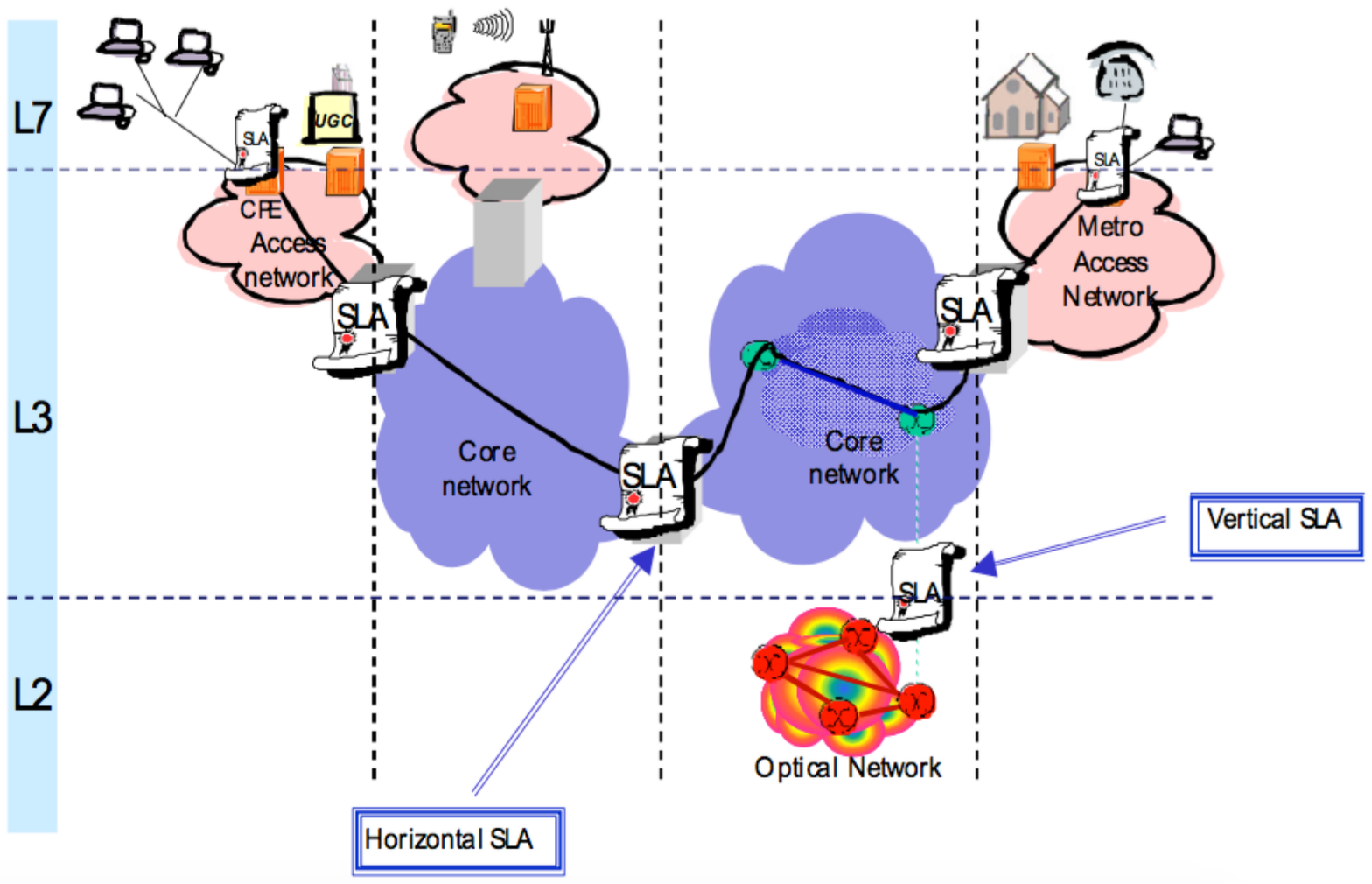}
\caption{A network topology including horizontal and vertical SLAs, adapted from~\cite{Marilly2002}}
\label{HvsVSLA}
\end{figure}

An SLA defines the guaranteed level of a service property such as response-time, availability and consequent actions in case of non-compliance situations, including liability as well as compensations issues.\footnote{An example of a paper based and a traditional SLA contract is available at \url{http://www.slatemplate.com/ServiceLevelAgreementTemplate.pdf} }~Based on the topological architecture of the network, an SLA contract can be categorised into horizontal and vertical SLA. Since the classification of SLAs in terms of horizontal and vertical categories depends on the network topological architecture, which is out of this article's scope, both categories are introduced just as follows. A horizontal SLA is an agreement between two service-providers existing at the same architectural layer (as an instance two Internet Protocol (IP) domains \cite{Martin2002}\cite{DANTONIO2004} or two domains of Optical Transport Network (OTN)~\cite{NAFARIEH2013}). On the other hand, a vertical SLA is an agreement between two individual providers at two various architectural layers (for instance between an optical network and the core MPLS network). A network topology diagram including potential SLA agreements among different architectural layers is shown in Fig.~\ref{HvsVSLA} to illustrate the concept of horizontal and vertical SLAs in a real scenario~\cite{Marilly2002}.

Other than the SLAs' network topological architecture, an end-to-end solution for management of SLA is required to define services, parameters of Service Level Specifications (SLS) and a classification of the services. The focus on the level of service instead of a network level enables the definition of SLA, services and/or Quality of Service (QoS) independently from the underlying network's technology. A service has to be defined without ambiguity utilising SLS and three information types must be described: i) The QoX metrics as well as their corresponding thresholds; ii) A method of service performance measurement; iii) Service schedule. QoX represents  different quality requirements such as QoS, Quality of Transmission (QoT), Grade of  Service (GoS), Quality of Resilience (QoR), Quality of Energy (QoEn), Quality of Knowledge (QoK) and Quality of Information (QoI) \cite{german2016}, and the mentioned quality parameters will be discussed in Section~\ref{sec3}.

\subsection{SLA actors and elements}

\subsubsection{Actors}

A typical SLA involves two entities such as a contract either between \textit{End-User (EU)} and \textit{Service Provider (SP)}, or between SP and \textit{Infrastructure Provider (InP)}. In general terms, the complete scenario includes the three mentioned actors. The term SP is referring to corporations that supply data and communication services to their customers. SPs may manage networks by themselves, or they may integrate the other SPs services to deliver an entire  service to their clients/customers~\cite{german2016}\cite{SLAbook}. The SP can operate in different business forms such as an Internet Service Provider (ISP),  a carrier, Application Service Provider (ASP) or an operator. The client and/or customer are referring to the end users that make use of all provided services via SP. In this survey, we use EU instead of customers onwards. InP is another SLA actor which is rarely found in the literature and mostly EU and SP are the actors that are the focused point in the literature. The ISP refers to the InP that provides physical resources, operational infrastructure and the computing services for development, deployment and management of the applications in enterprise class. For all involved actors in an SLA negotiation, a win-win situation can be defined as quality requirements that are satisfied for all actors, the EU is charged a fair price and the ISP and SP adequately remunerated. 

\subsubsection{Elements}

Table~\ref{elements} lists the SLA key elements along with a short description~\cite{SLAelements}. An SLA must be \textit{Specific} and detailed enough to define expectations for services and eliminate any confusion. The \textit{Comprehensiveness} is an essential element of the agreement and the SLA contract must cover all provided services by the SP and all possible contractual obligations for all actors involved. Moreover, the SLA should be directly related to the service to be offered and it must be \textit{Relevant} to evaluating performance against that goal. In the agreement, unrealistic goals can demotivate the customers and non-delivery will only lead to failures on agreed terms. Therefore, the expectations set must be \textit{Realistic}. By keeping the language simple and \textit{Non-technical}, for reference of EUs, the contract would be easily understandable. The responsibility should be clearly defined as a set of \textit{Division of work} in the agreement. The SLA must contain a \textit{Time frame} against which the service will be delivered. The \textit{Escalation Metrics} must be clearly defined. Once the actor enters into the agreement, the client must be aware whom to refer in case the services were not rendered properly. Once all elements are considered in the agreement, the agreement document must be the \textit{Authoritative} document binding all actors.

\begin{table}[!htb]
\renewcommand{\arraystretch}{1.3}
\caption{The best practices' elements to draft a comprehensive and reliable agreement}

\label{table_example}
\centering
\footnotesize
\label{elements}
\begin{tabular}{l|p{5cm}}
\hline \hline
Element & Description \\
\hline \hline
Authoritative & The SLA must be an authoritative document binding both parties.\\
Comprehensive & It should cover all services and all possible contractual obligations.\\
Division of work & The responsibility should be clearly defined.\\
Escalation matrix & The escalation matrix is a mapping of whom should be contacted under a particular set of conditions. For example, SLA must provide an escalation matrix for handling any issues of provided services by the SP and it must be clearly defined. \\
Measurable & There must be a way to track actual performance against the promised SLA.\\
Non-technical  & The SLA document must keep the language simple for reference of non-technical people.\\
Quantified & Deliverables should be quantified enabling to be measured.\\
Realistic & The expectations set by the SLA must be realistic.\\
Relevant & The SLA must be directly related to the service to be offered and delivered.\\
Specific & The SLA must be specific and detailed enough to define expectations for services.\\
Time frames & The SLA must contain a time frame against which the service will be delivered.\\
\hline \hline
\end{tabular}
\end{table}

\subsection{Literature classification}
\begin{figure}[h]
\centering
\includegraphics[draft=false,width=3.1 in]{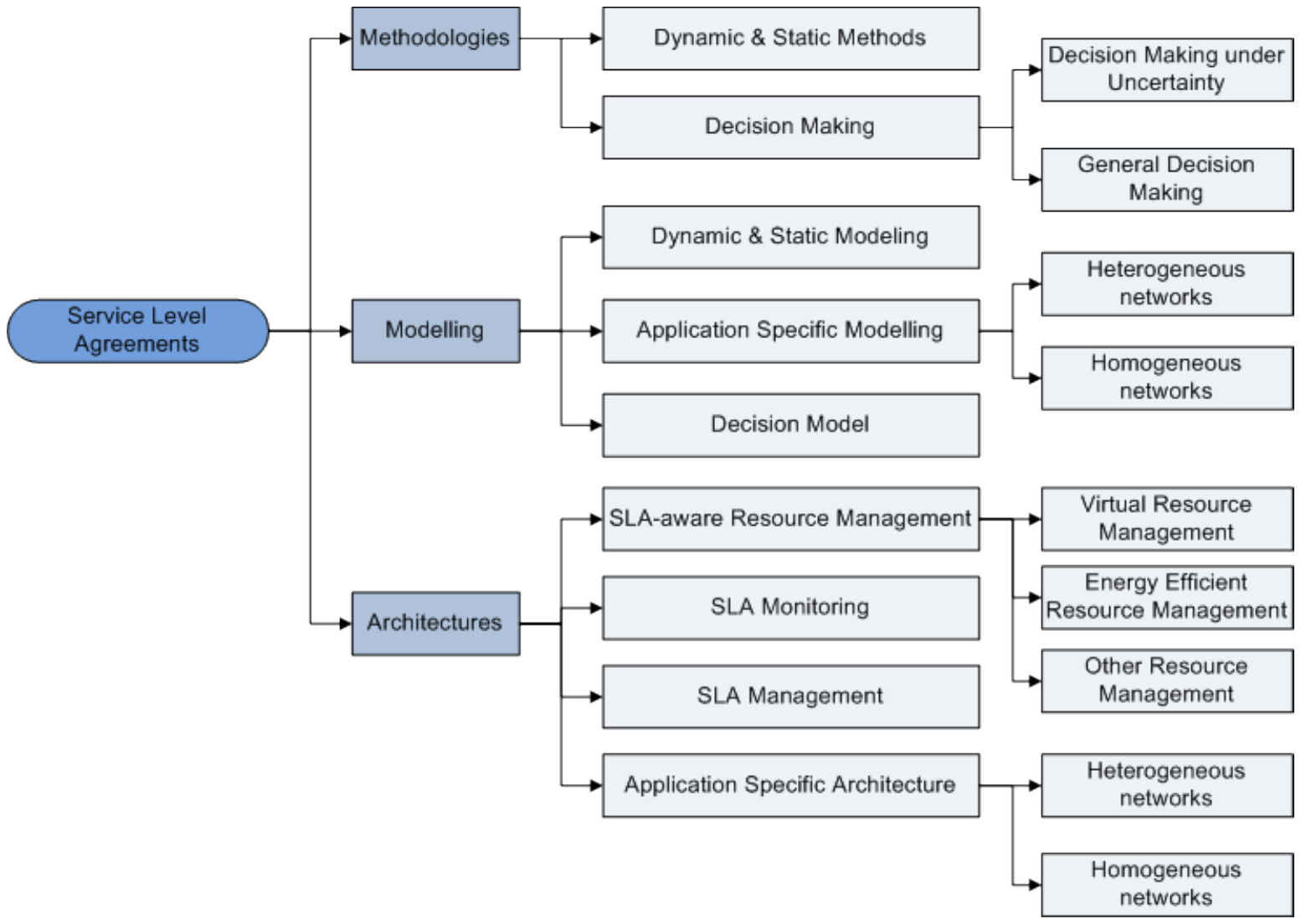}
\caption{Proposed classification of SLA literature}
\label{literatureclassification}
\end{figure}

To write this survey, more than 120 research articles and technical reports have been reviewed. Based on collected information on SLA, we defined a comprehensive conceptual-map (see  Appendix~\ref{app1}) that served as the basis for a structured-classification of the SLA literature. The literature on SLAs is in very various categories; organising and structuring the relevant works in a systematic way is not a trivial task. The proposed classification scheme for SLA literature is illustrated in Fig.~\ref{literatureclassification}.

We identify three categories: methodologies, modelling and architectures for service level agreements. This classification allows grouping the reviewed articles under common umbrella and enables to review three perspectives separately. Within the first category, methodologies for SLA, two subcategories have been proposed: dynamic and static methods as well as decision-making methodologies for SLAs. In the second subcategory, two additional subcategories have been identified: decision-making under uncertainty and general decision-making methods.

In the second category of the proposed classification, we identified three subcategories: dynamic and static modelling; application specific modelling; decision models for SLAs. Under the application specific modelling subcategory, we can extend our discussion in two different network classes: heterogeneous networks and homogeneous networks. Using this classification, it is possible to accommodate different network technologies under these two groups. 

In the architectures category for SLAs, we identified four subcategories. The first subcategory is SLA-aware resource management and it has three subcategories: virtual resource management, energy-efficient resource management and other resource management. SLA monitoring and SLA management have been also identified as subcategories for SLA architectures. Similar to SLA modelling we have application specific architecture for service level agreements. This part of proposed classification focuses on specific architecture proposed and/or deployed over heterogeneous networks or homogeneous networks. 

\subsection{Paper contribution}
This survey identifies the current SLA research status and provides a better understanding of existing methods and open problems in this research domain. The identification of research gaps and future research directions points to how to utilise SLA elements to develop a generic methodology for SLAs establishment, deployment and management that will lead to a win-win situation for all involved actors. Although there are many surveys about SLAs, so far they are mostly focused on particular services such as SLA in cloud computing. This paper is organised in a way that the reader can access a variety of proposed SLA methods and models addressed to different applications domains along with a detailed comparison. It needs to be emphasised that auction-based kind of SLAs are out of the scope of this survey.

\subsection{Paper organisation and reading-map}
The remainder of this survey is organised as follows: Section~\ref{sec2} introduces SLA characteristics and Service Level Objectives (SLO). Section~\ref{sec3} discusses service quality categories and quality-based service description. In that section we discuss four quality categories namely performance, configuration, data and security along with their quality parameters. Section~\ref{sec4} reviews the SLA methodologies proposed so far with focus on their behaviour and decision-making strategies while Section~\ref{sec5} reviews them from a modelling and architecture perspective. Section~\ref{sec6} focuses on the open problems and future research directions. Finally Section~\ref{sec7} concludes the survey. 

An “a la carte” approach can be followed to study this survey and Table~\ref{readingmaptab} provides a reading-guideline.  The readers with main interests in the detailed methodology and decision-making may focus their reading on Sections~\ref{sec1},~\ref{sec4},~\ref{sec6} and~\ref{sec7}. When it comes to those mainly interested in the detailed SLA Modelling aspects, they may read Sections~\ref{sec1},~\ref{sec2},~\ref{sec3},~\ref{sec5.1},~\ref{sec6} and~\ref{sec7}. The readers interested in SLA Architecture aspects can follow the Sections~\ref{sec1},~\ref{sec2},~\ref{sec3},~\ref{sec5.2},~\ref{sec6} and~\ref{sec7}. Finally, we recommend Sections~\ref{sec1},~\ref{sec2},~\ref{sec3},~\ref{sec6} and~\ref{sec7} to the group of readers who are interested in getting a very high-level overview of SLAs including the characteristics and objectives of SLA as well as open problems and future research directions in this domain.

\begin{table}[h]
\caption{A reading-guideline}
\centering
\footnotesize
\label{readingmaptab}
\begin{tabular}{|c|c|c|c|c|}
\hline \hline
& \multicolumn{4}{c|}{Readers interested in SLA}\\ \cline{2-5}
Sections & Basics & Methods & Models & Architecture\\
\hline \hline
1 &  $\checkmark$ & $\checkmark$ & $\checkmark$ & $\checkmark$ \\
2  & \checkmark & X  & \checkmark & \checkmark \\
3  & \checkmark & X & \checkmark & \checkmark \\
4  & X & \checkmark & X  & X\\

5.1 & X & X & \checkmark & X \\
5.2 & X & X & X  & \checkmark \\
 
6  & \checkmark & \checkmark & \checkmark & \checkmark \\

7 & \checkmark & \checkmark & \checkmark & \checkmark \\
\hline \hline
\end{tabular}
\end{table}


\section{SLA Characteristics and Objectives}\label{sec2}

In this section SLA characteristics and a key element named Service Level Objective (SLO) are explained. SLOs are specific and measurable characteristics of SLAs.

\subsection{SLA Characteristics}

The characteristics of SLAs can be identified as foundation, change and governance characteristics~\cite{Goo2009}. Foundation Characteristics (FC) of SLA includes provisions which specify the key agreements and principles among actors, the SP and their responsibilities and roles, and the expected service performance levels~\cite{Goo2009}. The aim behind the SLA agreements under FC is to publicise the normal convictions shared by the two associations with the goal that their Information and Communication Technology (ICT) outsourcing relationship could construct shared objectives and a general responsibility toward the outsourcing relationship~\cite{Choudhury2003}. By characterising the goals and aim of the relationship,  the objectives that at first drove the development of the relation can be at least understood partially  and shared by a group of decision-makers and also individual employees who inherit the relationship~\cite{Choudhury2003}\cite{Koh2004}. Moreover, clear procedures of conduct can be set by these provisions  through defining the responsibilities and roles  of the diverse actors involved in the SLA.

Legally binding expression relates with SLAs Change Characteristics (CC) including provisions concerning processes for resolving prospective demands unpredictable outcomes, processes for developing predicted contingencies as well as transformations, operations for recommending new innovations coordinated with motivating force designs, and processes of efficient adjustments and feedback in the bilateral contract~\cite{Goo2009}. The mentioned provisions are grouped under the change characteristics of a bilateral agreement. Moreover, these provisions over-attempt to make the ground procedures and rules for dealing with prospective contingencies. These provisions are depend on favourable outcomes as the environment of ICT develops quickly and the condition of business frequently needs quick reaction from the providers to provide new services or amend the running services ~\cite{Kirsch1997}. Although a comprehensive contracting is not a practical option, because of limited rationality, limited but intentional rationality is translated into incomplete but farsighted contract. Indeed, previous studies in Information Technology (IT) outsourcing have shown the impacts of evolving specifications and investigation the possibility for extremely unstructured and/or uncertain tasks~\cite{Choudhury2003}.

Contractual qualifications connected with Governance Characteristics (GC) of SLAs characterise the ways to keep the relationships through clear measurements statement, incentives and penalty, options for termination and accountabilities, and well-defined processes for documented communication as well as processes for resolving and recognising of potential disputes~\cite{Goo2009}. In this way, the contractual bases underlying governance specifications set administrative procedures to continuous assess  of  the value which the correlation is producing for the diverse stakeholders and to guarantee that the relationship remains on course. Both financial analysts and authoritative scholars similarly support the spirit of the GC. For instance, it is proposed that rewards and results be connected so as to prevail with regards to overseeing between hierarchical connections (e.g., ~\cite{Kirsch1997}).

\subsection{Service Level Objectives (SLO)}

An SLO is a key element of SLA among its entities. SLOs are agreed as a means of performance measuring of the provider and simultaneously they are outlined as a way to avoid disputes between the two actors based on misunderstanding. The establishment of a reliable, safe and QoE-aware computer networking requires a set of services that goes beyond pure networking services. Therefore, in the paper this broader set of services will be taken into account and for each SLO the related services domains will be indicated. There is often times confusion in the use of SLO and SLA. SLOs are particular and measurable characteristics of the SLA like throughput, response time, availability, or quality~\cite{Walter2017}. Apart of SLO application domain, we can categorise the objectives in four categories, namely: i) performance service level objectives; ii) security service level objectives; iii) data management service level objectives; iv) personal data protection service level objectives~\cite{slo2014}\cite{Baset2016}.

\subsubsection{Performance service level objectives}

The performance SLOs categories are given in Table~\ref{PerformanceSLO}. For each category of performance SLO, relevant SLOs are given with their description. This table covers i) Availability, ii) Response time, iii) Capacity, iv) Capability indicators, v) Support and vi) Reversibility SLO categories.

\subsubsection{Security service level objectives}

Security SLOs can be classified into eight SLO categories named: i) Reliability, ii) Authentication, iii) Cryptography, iv) Security, v) Logging, vi) Auditing, vii) Vulnerability and viii) Service change.  For each security SLO category relevant SLOs are given and explained briefly in Table~\ref{SecuritySLO}.

\subsubsection{Data management service level objectives}

Data management is one set of SLOs categories that covers data relevant categories. As an instance, data classification defines two service oriented relevant SLOs. Table~\ref{datamanagementSLO} also addresses the other SLO categories: data mirroring, backup, life-cycle and its portability.

\subsubsection{Personal data protection service level objectives}

Data protection and especially personal data protection are specific and measurable categories of SLO. Table~\ref{PersonalSLO} addresses eight individual SLO categories along with their relevant SLOs.

\begingroup
\onecolumn

\vspace{1.5cm}
\begin{scriptsize}
\begin{longtable}{c|p{3cm}|p{3cm}|p{6cm}}
\caption{Performance service level objectives}
\label{PerformanceSLO}\\

\hline \hline 
SLO Category & Relevant SLO & Service Domain &  Description\\
\hline \hline
\endfirsthead

\multicolumn{4}{c}%
{{\bfseries \tablename\ \thetable{} -- continued from previous page}} \\
SLO Category & Relevant SLO & Service Domain &  Description\\ \hline \hline
\endhead

\hline \multicolumn{4}{|r|}{{Continued on next page}} \\ \hline 
\endfoot

\hline \hline
\endlastfoot

 \multirow{6}{*}{Availability}  & Uptime level (availability) & Connectivity, Cloud Computing, VPN\footnote{Virtual private network}, SDN\footnote{Software-defined networking}, SFC\footnote{Service function chaining, also known as network service chaining} &  Shows the availability of service a certain period of the time, over the aggregated feasible available time (in percentage).\\ \cline{2-4}
 
& Successful requests percentage &  Connectivity, Cloud Computing, VPN, SDN, SFC &  Indicates the number of error-free requests processed by the service upon the collected number of submitted requests (in percentage). \\ \cline{2-4}

& Timely provisioning service requests rate &  Connectivity, Cloud Computing, VPN, SDN, SFC &  Indicates the number of provisioning service requests accomplished in a certain period of time over the total number of provisioning requests service (in percentage).\\ \hline
\multirow{4}{*}{Response time} & Average/Mean response time &  Connectivity, Cloud Computing, VPN, SDN, SFC &  The statistical-mean upon a set of observed response time service set for a particular type of request.\\ \cline{2-4}

& Maximum response-time &  Connectivity, Cloud Computing, VPN, SDN, SFC &  The maximum response-time goal for a particular and specific type of request.\\ \hline

\multirow{9}{*}{Capacity} &  Maximum resource capacity & Connectivity, Cloud Computing, SDN, SFC &  The highest and available amount of an allocated resource to an instance of the service for an special service client. Example of resource includes data storage, memory, number of CPU cores, and etc.\\ \cline{2-4}

& Simultaneous service users number &  Connectivity, Cloud Computing, SDN, SFC &  The maximum number of individuals that can be utilising the service at the same time. \\ \cline{2-4}

& Simultaneous connections  number &  Connectivity, Cloud Computing, SDN, SFC &  The maximum number of individual connections to the service at one time.\\ \cline{2-4}

& Service throughput &  Connectivity, Cloud Computing, VPN, SDN, SFC &  The minimum number of specific requests that can be processed by the service in an offered period of time like the number of requests per minute.\\ \hline
\multirow{2}{*}{Capability indicators} & External connectivity &  Connectivity, Cloud Computing, VPN, SDN, SFC &  Explains the service capabilities to connect to external systems and/or services. \\ \hline
\multirow{6}{*}{Support} & Support hours &  Cloud Computing, VPN &  Indicates the period of time in hours that SP provides an EU support interface that allows the requests  and general inquiries from the service client. \\ \cline{2-4}

& Support responsiveness &  Computer Networking, Cloud Computing &  The maximum period of time that the SP will take to acknowledge a service client request or inquiry. \\ \cline{2-4}

& Resolution time &  Computer Networking, Cloud Computing &  The target resolution-time for client requests, in other words, the time taken to accomplish any essential actions as a result of the request. \\ \hline
\multirow{7}{*}{Reversibility} & Period of data retrieval  & Cloud Computing &  Indicates the period of time that the EU can retrieve a copy of their service client's data from the service.\\ \cline{2-4}

& Data retention period &  Connectivity, Cloud Computing &  Indicates the period of time that the SP will maintain backup copies of the service client's data within the process of termination. \\ \cline{2-4}

& Residual data retention &  Connectivity, Cloud Computing &  Elucidate a description of any data related to the service client which is maintained since the termination process finished. \\

\end{longtable}
\end{scriptsize}

\vspace{1.2cm}
\begin{scriptsize}
\begin{longtable}{c|p{3cm}|p{3cm}|p{6cm}}
\caption{Security service level objectives}
\label{SecuritySLO}\\

\hline \hline 
SLO Category & Relevant SLO & Service Domain & Description\\
\hline \hline
\endfirsthead

\multicolumn{4}{c}%
{{\bfseries \tablename\ \thetable{} -- continued from previous page}} \\
SLO Category & Relevant SLO & Service Domain & Description\\ \hline \hline
\endhead

\hline \multicolumn{4}{|r|}{{Continued on next page}} \\ \hline 
\endfoot

\hline \hline
\endlastfoot

\multirow{4}{*}{Reliability} & Service reliability & Connectivity, Cloud Computing, VPN, SDN, SFC & Explains the service ability to accomplish its tasks accurately without any failure over some determined period of time. \\ \cline{2-4}

& Redundancy level & Connectivity, Cloud Computing, SFC & Describes the level of service supply chain redundancy.  \\ \hline
\multirow{11}{*}{Authentication} & User authentication and level of identity assurance & Connectivity, Cloud Computing, VPN, SDN, SFC & Represents the Level of Assurance (LoA) of the mechanism that is used to authenticate an EU to get access to a resource.\\ \cline{2-4}

& Authentication & Connectivity, Cloud Computing, VPN, SDN, SFC & Represents the available authentication-mechanisms supported by the Configuration Service Provider (CSP) on the offered services. \\ \cline{2-4}

& Revoke user access mean time & Connectivity, Cloud Computing, VPN, SDN, SFC & The average time required to cancel the users’ access to an specific service on request over a particular duration. \\ \cline{2-4}

& Storage protection & Computer Networking, Cloud Computing, VPN & Refers to the mechanisms utilised to protect a service-user access credentials. \\ \cline{2-4}

& 3rd-party authentication &  Connectivity, Cloud Computing & Identifies if 3rd-party authentication is supported by the service and also it describes which technology/technologies can be utilised for 3rd-party authentication. \\ \hline
\multirow{2}{*}{Cryptography} & Cryptographic brute force resistance &  Cloud Computing, VPN, SDN, SFC & Expresses the cryptographic protection strength applied to a certain resource based on its key length. \\ \cline{2-4}

& Key Access Control Policy (KACP) & Cloud Computing, VPN, SDN, SFC & Explains how a cryptographic key is protected strongly from access, when it is utilised to come up with security to a particular service. \\ \cline{2-4}

& Level of cryptographic hardware module protection & Cloud Computing, VPN, SDN, SFC & Refers to the protection level which is given to cryptographic processes through the use of cryptographic hardware modules in the service . \\ \hline
\multirow{8}{*}{Security} & Timely incident reports rate & Cloud Computing, VPN, SDN, SFC & Refers to the outlined incidents to the services that  are reported to the clients in a convenient time (in percentage). \\ \cline{2-4}

& Timely incident responses rate & Cloud Computing, VPN, SDN, SFC & Indicates the outlined incidents which are acknowledged and assessed by the SP in a convenient time (in percentage).\\ \cline{2-4}

& Timely incident resolutions rate & Cloud Computing, VPN, SDN, SFC & Explains the percentage of defined incidents on the service which are resolved after discovery and within a predefined time limit.\\ \hline
\multirow{3}{*}{Logging} & Logging parameters & Connectivity, Cloud Computing, VPN, SDN, SFC & Indicates the parameters which are obtained in the log files of a particular service. \\ \cline{2-4}

& Log access availability & Cloud Computing, VPN, SDN, SFC & Refers to the level of the service customer access to the log file entries. \\ \cline{2-4}

& Logs retention period & Cloud Computing, VPN, SDN, SFC & Indicates the duration that logs are available for further analysis. \\ \hline
\multirow{1}{*}{Auditing} &  Certifications applicable & Cloud Computing, VPN, SDN, SFC & Indicates a group of certifications maintained for a particular service by the SP.\\ \hline
\multirow{6}{*}{Vulnerability} & Timely vulnerability corrections rate & Cloud Computing, VPN, SDN, SFC & Indicates the number of vulnerability corrections performed by the SP (in percentage). \\ \cline{2-4}

& Timely vulnerability reports rate & Connectivity, Cloud Computing, VPN, SDN, SFC & Indicates the number of vulnerability reports generated the SP to the service user (in percentage). \\ \cline{2-4}

& Vulnerability corrections report & Cloud Computing, VPN, SDN, SFC & Refers to the mechanism that the SP informs the customer of vulnerability corrections applied to the SP's systems. \\ \hline
\multirow{5}{*}{Service change} &  Service change reporting & Cloud Computing, VPN, SDN, SFC & Refers to the sort of changes (such as functional changes or SLA change), mechanisms and a certain duration for SP to inform service clients of planned changes to the service. \\ \cline{2-4}

& Timely service change notifications percentage & Cloud Computing, VPN, SDN, SFC & The aggregate number of notifications for any change made during a particular period of the time over the whole number of notifications of change, in percentage. \\

\end{longtable}
\end{scriptsize}


\vspace{1.2cm}

\begin{scriptsize}
\begin{longtable}{c|p{3cm}|p{3cm}|p{6cm}}
\caption{Data management service level objectives}
\label{datamanagementSLO}\\

\hline \hline 
SLO Category & Relevant SLO & Service Domain & Description\\
\hline \hline
\endfirsthead

\multicolumn{4}{c}%
{{\bfseries \tablename\ \thetable{} -- continued from previous page}} \\
SLO Category & Relevant SLO & Service Domain & Description\\ \hline \hline
\endhead

\hline \multicolumn{4}{|r|}{{Continued on next page}} \\ \hline 
\endfoot

\hline \hline
\endlastfoot

\multirow{3}{*}{Data-classification} & Service client data use by SP & Cloud Computing, SDN, SFC & Indicates the stated policy for any service client data use. \\ \cline{2-4}

& Service derived data-use & Cloud Computing & Explains what derived data is produced by the SP from service client data. \\ \hline
\multirow{13}{*}{Data mirroring/backup} & Data mirroring latency & Cloud Computing & Indicates the difference between the time data is located on mirrored-storage and the time which the same data is located on primary-storage.\\ \cline{2-4}

& Data backup method & Computer Networking, Cloud Computing & Indicates a list of methodologies that are used to make a backup service client's data.\\ \cline{2-4}

& Frequency of data backup & Computer Networking, Cloud Computing & Explains the time interval in between full backups of service client's data.\\ \cline{2-4}

& Backup retention time & Cloud Computing & Indicates how long a designated backup exists to be utilised for data restoration. \\ \cline{2-4}

& Backup generations & Computer Networking, Cloud Computing & Explains the total number of backup generations existing to be utilised for data restoration..\\ \cline{2-4}

& Maximum data restoration-time & Computer Networking, Cloud Computing & Explains the committed time that is taken to restore service customer data from a backup.\\ \cline{2-4}

& Successful data restorations rate & Computer Networking, Cloud Computing & Indicates the committed  data restorations success rate, expressed as the total number of error-free data restorations carried out for the clients upon the aggregate number of data restorations (in percentage).\\ \hline

\multirow{7}{*}{Data life-cycle} &  Type of data deletion & Cloud Computing & Refers to the quality of data elimination, ranging from the weak elimination that only remove the reference to the data, to strong sanitisation methods to ensure that removed data cannot be easily recovered.\\ \cline{2-4}

& Timely effective deletions rate & Cloud Computing & Describes the total number of the requests for data elimination by client accomplished during a pre-defined period of time limit over the whole number of requests for data deletion (in percentage).\\ \cline{2-4}

& Tested storage retrieve-ability rate & Cloud Computing & Indicates the amount of verified clients data to be recoverable within the measurement period, right after the data has been removed. \\ \hline

\multirow{7}{*}{Data portability} & Data portability format & Cloud Computing, SDN, SFC & Refers to the e-format(s) that service client data may get approached by or transferred to  the service. \\ \cline{2-4}

& Data portability interface & Cloud Computing, SDN, SFC & Refers to the methods that can be utilised to transfer customers' data to the service or from the service.\\ \cline{2-4}

& Data transfer rate & Cloud Computing, VPN, SDN, SFC & Indicates the minimum rate of service customer data that can be transferred from and/or to the service through using the method(s) offered in the data interface. \\

\end{longtable}
\end{scriptsize}

\vspace{1.2cm}

\begin{scriptsize}
\begin{longtable}{c|p{3cm}|p{3cm}|p{6cm}}
\caption{Personal data protection service level objectives}
\label{PersonalSLO}\\

\hline \hline 
SLO Category & Relevant SLO & Service Domain & Description\\
\hline \hline
\endfirsthead

\multicolumn{4}{c}%
{{\bfseries \tablename\ \thetable{} -- continued from previous page}} \\
SLO Category & Relevant SLO & Service Domain & Description\\ \hline \hline
\endhead

\hline \multicolumn{4}{|r|}{{Continued on next page}} \\ \hline 
\endfoot

\hline \hline
\endlastfoot

\multirow{3}{*}{Conduct codes} & Applicable data protection codes of conduct, certifications, standards & Cloud Computing, VPN, SDN, SFC & Refers to a list of certification mechanisms and methods, standards and the data protection codes of conduct  that the service agreed with.\\ \hline
\multirow{2}{*}{Purpose specification} & Processing purposes & Cloud Computing, SDN, SFC & Indicates the list of purposes of processing which are beyond the requests by the clients acting as a controller. \\ \hline
\multirow{5}{*}{Data minimisation}& Period of temporary data retention & Cloud Computing, SDN, SFC & Indicates the utmost duration in which provisional data is maintained after recognition that the provisional data is unused. \\ \cline{2-4}

& Period of service customer data retention & Cloud Computing, SDN, SFC & Describes the utmost duration that a service client data is maintained right before destruction by SP and after acknowledgement of a request to erase the data or contract termination.\\ \hline
\multirow{6}{*}{Use limitation} & Indicates number of customer data law enforcement disclosures & Cloud Computing & Indicates the number of private data  disclosures to law  enforcement authorities  over a  predefined period of time. \\ \cline{2-4}

& Personal data disclosure notifications & Cloud Computing & Describes the number of personal and private data  disclosures to  law enforcement  authorities  notified to the clients over a  certain period of time. \\ \hline
\multirow{6}{*}{Transparency} & Contractors and subcontractors lists & Cloud Computing & Indicates the SP's contractors and subcontractors that participate in the processing of the client's private data service.\\ \cline{2-4}

& Special data categories & Cloud Computing & Indicates the list of the particular categories of personal data such as financial and health related data or sensitive data.\\ \hline
\multirow{6}{*}{Accountability} & Documentation & Cloud Computing, VPN, SDN, SFC & Describes a list of the documents which is made available by the SP  in order to demonstrate the admission to the data protection obligations and needs.\\ \cline{2-4}

& Personal data breach policy & Cloud Computing, VPN, SDN, SFC & Refers to the policy of the SP regarding data breach. \\ \hline

\multirow{5}{*}{Data location} & Data geolocation list & Cloud Computing & Determines the geographical position where the clients' data processed and also stored by the SP. \\ \cline{2-4}

& Selection data geolocation & Cloud Computing & Describes whether clients can choose an specific geographical site to store the service client data.\\ \hline
\multirow{3}{*}{Intervene-ability} & Access request response time & Cloud Computing, SDN, SFC & Describes the period of time in which the SP must communicate the essential information to permit the client to respond to access-requests via the data subjects. \\

\end{longtable}
\end{scriptsize}

\endgroup
\twocolumn

\section{Service Quality Categories \& QSD}\label{sec3}

Although all quality requirement parameters (QoX) play a significant role in SLA, within an agreed/standard service life-cycle, Fig.~\ref{ServiceLifeCycle} (adapted from~\cite{Kritikos2013}) does not show QoX and QoS is the only one that is taken into account, what is quite insufficient nowadays. Quality documents are necessary for each and every one of the activities of the reference service life-cycle and their importance is explained below:

\begin{figure}[h]
\includegraphics[draft=false,width=3in,center]{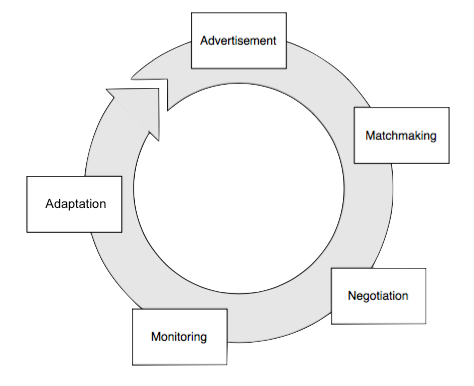}
\caption{Service life-cycle, adapted from~\cite{Kritikos2013}}
\label{ServiceLifeCycle}
\end{figure}

\begin{itemize}

\item \textit{Advertisement:} EUs and SPs publish and/or exchange the quality requests and offers, respectively. Both quality documents are named as ''Quality-Based Service Descriptions (QSDs)".

\item \textit{Matchmaking:} QSDs are matched up in order to determine if offers are capable to support the requirements of EUs. The consequence is that the advertised functionally equivalent services are filtered and selected based on their ability to satisfy the EU quality requirements afterwards.

\item \textit{Negotiation:} QSDs or SLA templates are interchanged between SP and EU as well as between InP and SP. The feasible agreement between the actors involved leads to the definition of an SLA.

\item \textit{Monitoring/Assessment:} the SLA is tracked in order to discover EU and/or SPs’ violations of their quality and functional terms and conditions. 

\item \textit{Adaptation:} The adaptation/recovery and proactive functions can be taken in case the SLA is unfulfilled. A feasible recovery action might need the matchmaking activity execution or a further SLA negotiation to discover an alternative service. It may also occur that an alert is sent to the evaluation component of the monitoring activity that pursues to execute.

\end{itemize}

QSDs are used in the first two and/or three service life-cycle activities, while SLAs are utilised in the last three service life-cycle activities. Therefore, there is not any standard and uniform document for the quality to be utilised among entire activities mentioned in the service life-cycle and this is a main drawback which is time consuming, as transformations of the documents must take place to another format from an origin template.

Although several different QoS categories and attributes can be found in different research reports, it is possible to extract few of them that appear most often and can be considered as essential QoS categories and attributes, respectively. The other attributes are mostly context-dependent (i.e., quite particular) or capture secondary features, as they appear very seldom in the proposed Services Quality Managements (SQMs) so far. Therefore, the most reported QoS categories and attributes can be considered as the most important ones. 

As shown in Table~\ref{QSD}, throughput,  latency, processing time and response time are the attributes that often demonstrate the category of performance, that is available in the majority of SQM proposals. Although it is not a main focus of this survey, security is the other significant category that has three foremost attributes among others, namely; authorization, authentication, and non-repudiation, which are steadily existent in the SQMs. Reliability, accuracy as well as availability are the other three most significant attributes that are not categorised under a certain service quality category. 

Data quality features are seldom considered in SQMs. The only data quality feature, that is frequently considered in reported articles, is ''correctness". The accuracy attribute defined by various contributions refers to the correctness of the service and also of the data that it provides. Since the output of a service is often compound of information, data quality may considered as a part of the QoS, and it can drive thoroughly the analysis of the provided output as well as necessary input. Data quality is a multidimensional concept that defines the suitability of the utilised data for the specific service in which they are involved~\cite{Kritikos2013}\cite{Kritikos2009}. The utmost significant and representative attributes of data quality, which should be part of a SQM are accuracy, completeness, consistency, and timeliness. Among the reviewed articles, \cite{Sabata1997} is a quite old reference but it is the most cited article in research papers along the US patent (Pub. No.: WO2001099349) for the QoS specifications area. These attributes can be utilised by the service for the data correctness investigation purpose, the existence of inconsistent or missing values, and the information updateness.

Some of the proposed SQMs take into consideration specific network aspects. Usually, there is a network quality category in these SQMs that comprises the four most frequent attributes of network, namely: packet loss ratio, jitter, network delay and bandwidth. Table~\ref{QSD} lists the attributes in service quality categories along the list of references for published approaches. However, most of the attributes are addressed in~\cite{Kritikos2013}~comprehensively.

\begin{table}[!htbp]
\caption{Service quality categories and attributes in service quality categories' approaches (Adapted from \cite{Kritikos2013})}
\label{QSD}
\footnotesize
\begin{tabular}{c|p{2cm}|p{3cm}}

\hline \hline 
Category & Attributes & SQM approaches references \\ \hline \hline

\multirow{9}{*}{Performance} & Response time & \cite{Kritikos2013}\cite{Kritikos2009}\cite{Anbazhagan2002}\cite{Ran2003}\cite{Colombo2005}\cite{Oasis2005}\cite{Cappiello2006} \cite{Truong2006}\cite{Cappiello2009}\cite{Frutos2009}\cite{Mabrouk2009} \\ \cline{2-3}

 & Processing time & \cite{Kritikos2013}\cite{Kritikos2009}\cite{Anbazhagan2002}\cite{Cappiello2006}\cite{Truong2006}\cite{Cappiello2009}\cite{Liu2004} \\ \cline{2-3}
 
 &  Latency  & \cite{Kritikos2013}\cite{Kritikos2009}\cite{Anbazhagan2002}\cite{Ran2003}\cite{Colombo2005}\cite{Cappiello2006}\cite{Cappiello2009} \cite{Frutos2009}\\ \cline{2-3}

 &  Timeliness &  \cite{Kritikos2013}\cite{Colombo2005}\cite{Cappiello2006}\cite{Sabata1997}\\\cline{2-3}
 
 &  Precision  & \cite{Kritikos2013}\cite{Kritikos2009}\cite{Sabata1997}\\ \cline{2-3}
 &   Throughput & \cite{Kritikos2013}\cite{Anbazhagan2002}\cite{Ran2003}\cite{Colombo2005}\cite{Oasis2005}\cite{Cappiello2006}\cite{Cappiello2009} \cite{Frutos2009}\cite{Mabrouk2009}\\ \hline


\multirow{14}{*}{Security} & Authentication &  \cite{Kritikos2013}\cite{Kritikos2009}\cite{Ran2003}\cite{Oasis2005}\cite{Cappiello2006}\cite{Truong2006}\cite{Cappiello2009} \cite{Mabrouk2009}\\ \cline{2-3}

& Authorization &  \cite{Kritikos2013}\cite{Kritikos2009}\cite{Ran2003}\cite{Oasis2005}\cite{Truong2006}\cite{Cappiello2009}\cite{Frutos2009} \cite{Mabrouk2009}\\ \cline{2-3}

& Security Level & \cite{Kritikos2013}\cite{Kritikos2009}\cite{Truong2006}\cite{Sabata1997}\\\cline{2-3}

& Integrity & \cite{Kritikos2013}\cite{Kritikos2009}\cite{Anbazhagan2002}\cite{Ran2003}\cite{Oasis2005}\cite{Truong2006}\cite{Cappiello2009}\\ \cline{2-3}

& Confidentiality & \cite{Kritikos2013}\cite{Kritikos2009}\cite{Anbazhagan2002}\cite{Ran2003}\cite{Oasis2005}\cite{Cappiello2006}\cite{Truong2006} \cite{Cappiello2009}\cite{Frutos2009}\cite{Mabrouk2009}\cite{Sabata1997}\\ \cline{2-3}

& Accountability &  \cite{Kritikos2013}\cite{Kritikos2009}\cite{Ran2003}\cite{Truong2006}\cite{Cappiello2009}\cite{Frutos2009}\cite{Mabrouk2009}\\ \cline{2-3}

& Traceability & \cite{Kritikos2013}\cite{Kritikos2009}\cite{Ran2003}\cite{Oasis2005}\cite{Cappiello2009}\cite{Frutos2009}\cite{Mabrouk2009}\\ \cline{2-3}

& Non repudiation &  \cite{Kritikos2013}\cite{Kritikos2009}\cite{Ran2003}\cite{Oasis2005}\cite{Cappiello2009}\cite{Frutos2009}\cite{Mabrouk2009}\\ \cline{2-3}

& Data encryption & \cite{Kritikos2013}\cite{Kritikos2009}\cite{Ran2003}\cite{Oasis2005}\cite{Cappiello2006}\cite{Cappiello2009}\cite{Frutos2009} \cite{Mabrouk2009}\\ \cline{2-3}

& Isolation & \cite{Kritikos2013}\\ \hline


\multirow{4}{*}{Data} & Maturity/age & \cite{Kritikos2013}\cite{Truong2006}\cite{Cappiello2009} \\ \cline{2-3}

& Timeliness & \cite{Kritikos2013}\cite{Kritikos2009}\cite{Truong2006}\cite{Cappiello2009} \\\cline{2-3} 
 
& Reliability & \cite{Kritikos2013}\cite{Truong2006}\cite{Cappiello2009} \\\cline{2-3} 

&  Completeness & \cite{Kritikos2013}\cite{Kritikos2009}\cite{Truong2006}\cite{Cappiello2009} \\ \hline

 
\multirow{7}{*}{Configuration} & Virtual organisation & \cite{Kritikos2013}\cite{Cappiello2006}\cite{Truong2006}\\ \cline{2-3}

& Location & \cite{Kritikos2013}\cite{Truong2006}\\\cline{2-3} 

& Level of service & \cite{Kritikos2013}\cite{Cappiello2009}\cite{Sabata1997}\\ \cline{2-3} 

& Service version &   \cite{Kritikos2013}\cite{Truong2006} \\\cline{2-3}

& Supported standard & \cite{Kritikos2013}\cite{Kritikos2009}\cite{Anbazhagan2002}\cite{Ran2003}\cite{Colombo2005}\cite{Oasis2005}\cite{Truong2006} \cite{Cappiello2009}\cite{Mabrouk2009} \\ \hline


\multirow{28}{*}{Uncategorised} & Cost & \cite{Kritikos2013}\cite{Kritikos2009}\cite{Ran2003}\cite{Cappiello2006}\cite{Truong2006}\cite{Cappiello2009}\cite{Frutos2009} \cite{Liu2004}\cite{Sabata1997} \\ \cline{2-3}

& Availability & \cite{Kritikos2013}\cite{Kritikos2009}\cite{Anbazhagan2002}\cite{Ran2003}\cite{Colombo2005}\cite{Oasis2005}\cite{Cappiello2006} \cite{Truong2006}\cite{Cappiello2009}\cite{Frutos2009}\cite{Mabrouk2009} \\ \cline{2-3}

& Accessibility &  \cite{Kritikos2013}\cite{Kritikos2009}\cite{Anbazhagan2002}\cite{Oasis2005}\cite{Cappiello2006}\cite{Truong2006}\cite{Cappiello2009} \cite{Frutos2009}\cite{Mabrouk2009}\\ \cline{2-3}

& Accuracy &  \cite{Kritikos2013}\cite{Kritikos2009}\cite{Anbazhagan2002}\cite{Colombo2005}\cite{Cappiello2006}\cite{Truong2006}\cite{Cappiello2009} \cite{Frutos2009}\cite{Mabrouk2009}\cite{Sabata1997} \\ \cline{2-3}

& Reliability & \cite{Kritikos2013}\cite{Kritikos2009}\cite{Anbazhagan2002}\cite{Ran2003}\cite{Colombo2005}\cite{Cappiello2006}\cite{Truong2006} \cite{Cappiello2009}\cite{Frutos2009}\\ \cline{2-3}

& Capacity & \cite{Kritikos2013}\cite{Kritikos2009}\cite{Anbazhagan2002}\cite{Ran2003}\cite{Truong2006}\cite{Cappiello2009}\cite{Frutos2009}\\ \cline{2-3}

& Believability & \cite{Kritikos2013}\cite{Truong2006}\\ \cline{2-3}

& Maintainability & \cite{Kritikos2013}\cite{Kritikos2009}\cite{Colombo2005}\cite{Truong2006}\\ \cline{2-3}

& Relative importance & \cite{Kritikos2013}\cite{Sabata1997}\\ \cline{2-3}

& Complexity &  \cite{Kritikos2013}\cite{Colombo2005}\\\cline{2-3}

& Customer service &  \cite{Kritikos2013}\cite{Colombo2005} \\\cline{2-3}

& Dependability &  \cite{Kritikos2013}\cite{Kritikos2009}\cite{Anbazhagan2002}\cite{Colombo2005}\cite{Cappiello2006}\cite{Cappiello2009}\\ \cline{2-3}

& Stability &  \cite{Kritikos2013}\cite{Kritikos2009}\cite{Anbazhagan2002}\cite{Ran2003}\cite{Colombo2005}\cite{Oasis2005}\cite{Cappiello2009} \cite{Mabrouk2009} \\ \cline{2-3}

& Trust &  \cite{Kritikos2013}\cite{Colombo2005}\cite{Liu2004} \\ \cline{2-3}

& Understandability &  \cite{Kritikos2013}\cite{Colombo2005}\cite{Cappiello2009} \\\cline{2-3}

& Integrability &  \cite{Kritikos2013} \\ \cline{2-3}

& Interoperability &  \cite{Kritikos2013}\cite{Anbazhagan2002}\cite{Cappiello2006} \\ \cline{2-3}

& Resource efficiency &  \cite{Kritikos2013} \\ \cline{2-3}

& Re-usability &  \cite{Kritikos2013} \\ \cline{2-3}

& Scalability & \cite{Kritikos2013}\cite{Anbazhagan2002}\cite{Ran2003}\cite{Cappiello2006}\cite{Cappiello2009} \\ 
\hline\hline
\end{tabular}
\end{table}

\section{SLA Methodologies}\label{sec4}

The preparation and deployment of a robust method that combines both economical and technical aspects in order to offer a contract to satisfy all actors in an SLA scenario is an important task to be done. Hereafter, in this article, a contract that can satisfy EU, SP and SP requirements is named a \textit{win-win contract} and it may not be the purpose of all service providers. However the analysis and conclusions of this work can be applied to all contexts. Although some methods have been reported and published for SLAs, yet a lack of a comprehensive model makes this research domain open for further research. In this section we review available methods proposed so far and we categorise them in dynamic and static methods and also on decision-making based methods.\\

\subsection{Dynamic \& Static Methodologies}
There are significant differences between behaviours of run-time service as well as the expectations of quality stated in the contract with the clients due to the informal SLAs' nature. Managing SLAs in ICT service industries could be handled through either \textit{Static} or \textit{Dynamic} methods~\cite{HE2007}\cite{Dey2010}\cite{Benlarbi2006}. Although there are challenges in developing methods for static SLAs to consider all QoX parameters rather than choosing the only QoS, the development of dynamic methods is an even more critical issue for SPs. This is because of the emerging technologies as well as the continuous and frequent changes in service needs and techniques over the time. Any SLA  updating to meet any desired change in a service requires re-mapping and reformulation of all criteria. Recently, dynamic solutions are becoming increasingly commonplace as SLA methods and most of the research articles focus on recent technologies such as cloud computing. As an instance, W.~Halboob \textit{et al.}, published 23 contributed research articles in this context~\cite{Halboob2014}\cite{Halboob2015}.

\subsection{Decision-Making}

The services industry in an ICT domain has been arrived to be a dominating activity in most advanced industrialized economies. Generally, beyond 75\% of the labour force is engaged in the domain of services industry in the United States alone, with an overall production demonstrating about 70\% of whole industry output \cite{Dubina2017}\cite{Javier2017} and, simultaneously, ICT service industries/providers are growing significantly fast nowadays.

The effective and efficient management of service providing processes of decision-making within the bounds of ICT industry is absolutely important to any SP in practice. However, an efficient decision-making is even difficult and very complex to achieve due to the various quality requirements from EUs point of view. Particularly this is the case for the processes of complex decision-making, involved in the ICT services delivery, extremely uncertain and potentially volatile client requirements, complex models of SLA, and also the necessity to capture diverse feasible aspects of ICT services delivery such as QoX. We can divide decision-making processes into two: i) \textit{General decision-making} and ii) \textit{Decision-making under uncertainty} categories. Furthermore, the decision-making under uncertainty has two subcategories that are defined as: i) \textit{General decision-making under uncertainty} and ii) \textit{Specific decision-making under uncertainty}, and most of SLA decision-making among SP, EU and InP could be utilised and accommodate under these two subcategories.

\subsubsection{General Decision-Making}
Recently research about SLA and its management has been soared because of the importance of the SLAs in ICT industries. A lot of the work deals with automated and also technically oriented runtime intelligent SLA negotiation \cite{Yaqub2014}\cite{erban2010}\cite{Franke2016}. Although most of the recent works did not emphasised and concentrate on QoX, they highlighted a key challenge in this area as the prediction of cost and QoS, as to be able to make an optimal decision already in the negotiation phase~\cite{Franke2016}\cite{Tang2014}\cite{ydiao2014}.

Once negotiations are accomplished and the SLA is deployed, the SP should be able to implement provision according to different policies, and finding such policies so as to achieve a win-win deal is another important area of research. As an instance, SPs can allocate resources to guarantee SLA fulfilment (guaranteed enforcement), or exclusively allocate some resources per need (lazy enforcement)~\cite{Franke2016}\cite{Aib2007}. Moreover, the risk and its management are important phases on SLAs and it significantly effects on business due to service downtime. For example T. Setzer \textit{et al.}~\cite{Setzer2008}, proposed a model for optimal service-window scheduling to minimize the business impact, but unplanned outages are not considered in this research and it might not be able to achieve a win-win deal and negotiation. 

Another significant research domain in SLA risk-management is evaluation of the probability that SLOs specified in the SLA will not be met as reported in \cite{Snow2007}\cite{Snow2010}. Moreover, to develop such models, knowledge of the involved statistical distributions is crucial \cite{Franke2014}.

\subsubsection{General Decision-Making Under Uncertainty}

The general decision-making approaches under uncertainty concern various fields such as cognitive science, engineering, artificial intelligence, economics and many others. However, it is out of the scope of this paper to review the vast literature on decision-making theories under uncertainty, but in general, most of the proposed solutions are based on either decision-making under risk~\cite{Kahneman1974} or strict-penalty~\cite{HUCK1999}. There are some reviews about this topic such as~\cite{Starmer2000} and however being a bit old, it covers most of the fundamental required knowledge. In this area, a study of how the formulation of decision-making under uncertainty can induce deviations from maximizing expected monetary value found that even with small expected values differences, subjects on average did maximize expected value in their choices~\cite{HUCK1999}. Another study reported that maximization of expected value to be the best explanation of subject behaviour in so called \textit{duplex gambles} in the losing form (where the subject can lose but not win money), but not in the winning form (where the subject can win but not lose money)~\cite{DAVENPORT1973}. 

However, it must be noted that the purpose of this survey is not to contribute to the general descriptive question of decision-making under risk/strict-penalty and the purpose of this section is just to introduce SLA decision-making to be considered to use in the SLA methodology development.

\subsubsection{Specific/Particular Decision-Making Under Uncertainty}

Beyond the general studies of decision-making under uncertainty, also a number of more specific studies have been conducted to describe particular contexts or particular stakeholders such as entrepreneurs. One study addresses the decision-making of CEOs~\cite{List2011}. Their behaviour seems to deviate a lot from expected utility maximization. Another research, looking at entrepreneurs in China, concludes that even though the entrepreneurs are more willing to accept strategic uncertainty (related to competition and trust), they do not differ from the control group when it comes to non-strategic uncertainty, such as risk~\cite{Holm2013}.

These results are consistent with a study of entrepreneurs in Denmark, who do not seem to be more or less risk-averse than the total population, nonetheless they are rather optimistic regarding the chance of occurrence for the best consequence in lotteries with real money \cite{Andersen2014}. In \cite{LEFEBVRE2014}, M. Lefebvre \textit{et al,} report that the risk taking of executives under different incentive contracts and the impact of different incentive contracts (stock options vs. stock shares) for executives was investigated using students. The results indicate lots of excessive risk-taking, in particular risk-seeking behaviour in the face of small probability gains and large probability losses.

The comprehensive literature on decision-making under uncertainty in specific/particular contexts, some of which has been cited above, indicates that such contextual investigations are considered worthwhile. There is no simple and/or straightforward ways to generalise across contexts, instead, a piecemeal approach is obvious in the literature. It requires to be emphasised that the ICT SLA decision-making context has, so far, not been investigated.

\section{Modelling and Architecture}\label{sec5}

A successful SLA requires a comprehensive modelling and a robust architecture. The SLA model develops a sincere functional service description to permit for the expression of service quality guarantees as well as non-functional service properties. Furthermore, SLA comes up with the architecture along with an specification of the quality characteristics that the service will provide. This specification allows the architecture to pick out the service that best supports the system's quality attribute requirements. This section presents the research findings and reports on SLA modelling and architecture along with the monitoring which is one of the important objectives for SLA.

\subsection{SLA Modelling}
\label{sec5.1}
Yixin Diao \textit{et al.}, proposed a modelling framework that uses queueing-model-based approaches for estimation of the impact of SLAs on the delivery cost. Furthermore, they proposed a set of approximation techniques to address the complexity of service delivery and an optimization model to predict the delivery cost subject to service-level constraints and service stability conditions~\cite{ydiao2014}. Another recent work on this context introduced new SLA scenarios and considered new quality parameters in SLA modelling. Moreover, the use of SDN paradigm  has been proposed and discussed to implement the fulfilment of the SLAs established among the actors~\cite{german2016}.

The SLA negotiation is an essential mechanism to guarantee the performance of supplied service and to enhance the trust between EUs and SPs. In this area there are studies published that focus on the negotiation part of SLA. As an instance, in~\cite{ietf2000} an outline for the definition of a Service Level Specifications (SLSs) format, the semantics, and some early ideas on the requirements of negotiation of SLSs have been discussed. A mutual negotiation protocol applying an alternative offers model for resource allocation and scheduling in Grid-Federation as well as a multi-steps SLA negotiation, that includes the cloud SP selection and the negotiation with the chosen SP are addressed in \cite{iccke2013} and \cite{anithakumari2017} respectively. To develop a general framework for strategic negotiation of service level values under time limitations, the latest developments in agent research are studied in \cite{erban2010}. Another study proposed a methodology which aims to appraise the service  performance at early stages of the development process using simulation. The simulation data may be utilised first to negotiate the SLA preliminary performance of the service between SPs and EUs looking for a given service, and later to monitor it \cite{Sagbo2016}. In \cite{silaghi2012} a framework has been proposed for strategic negotiation of service level values under time limitations and exemplify the usage of the framework through extending the ''Bayesian learning agent" to cope with the limited duration of a negotiation session. G. C. Silaghi \textit{et al.}, claimed that opponent learning strategies are worth to be considered in open competitive computational grids, leading to the fair satisfaction of actors and an optimal allocation of resources. On the other hand, for intra- and inter-domain service level negotiation, an extension of the COPS protocol has been proposed and named as COPS-SLS \cite{nguyen2002}. 
 
Among the reviewed studies in SLA domain, there are articles concentrated on SDN based approaches \cite{german2016}\cite{arkham2017}\cite{son2017}\cite{STANIK2014} and few of them contributed with new concepts in the area \cite{german2016}. On the other hand, with the fast growth of some other technologies such as cloud computing, researchers became interested in making more investigation on  SLA negotiation, development and deployment on the cloud. As an instance, in \cite{Franke2016} authors tried to answer a research question whether enterprise ICT professionals could maximize expected value when procuring availability SLAs, and they tried to explore an optimal solution for it, from the business point of view.

In general, SLAs are important in cloud computing as they establish agreements between the cloud customers and their service suppliers, concerning the standard of the provided service. Many research articles reported their findings on SLA domains over cloud computing and we can categorise these articles in eight categories: i) SLA negotiation and monitoring; ii) Reviews on QoS aware SLA; iii) SLA decision-making; iv) Security in SLA; v) Specific theories based SLA; vi)   Energy-efficient SLA solutions; vii) Cloud based architecture and modelling and viii) SLA aware resource allocation. 
   
In the first category, multi-step negotiation of SLA, that includes the selection of cloud providers and the negotiation with the selected SP is introduced in~\cite{anithakumari2017}. However, \cite{Lawrence2011} tried to come out with an optimal solution for infrastructure service under the European project called OPTIMIS. Furthermore, negotiating frameworks for SLA of cloud-based services are proposed by \cite{Yaqub2014}\cite{erban2010}\cite{STANIK2014}\cite{ELAWADI2015}. In fact, SPs hesitate to negotiate and instead providers prefer to offer strictly binding SLAs. It is because of the economic risk assessment exposure which is a significant challenge. Therefore it requires some specific models to bring the risk-aware SLA solutions such as \cite{Hedwig2012}\cite{Mastroeni2011}\cite{Macias2014}. Although most of the articles considered QoS parameter to validate their findings, there are significantly small numbers of articles that rely on QoE instead \cite{varela2015}\cite{Frangoudis2014}\cite{Sackl2013}.
 
One of the main challenges in the monitoring is monitoring violations. This is because the existing computer networking platforms allow users to build distributed, large,  and complex applications. Thus, it is quite critical to develop SLA monitoring solutions prediction and prevention of SLA violations. Here are some examples: \cite{Tang2014}\cite{Chhetri2016}\cite{Shenoy2018}\cite{leitner2010}\cite{1eitner2013}\cite{1eitner20101}\cite{faniyi2012}\cite{falasi2013}\cite{Morshedlou2014}\cite{xia2008}.

In the review process we found few review articles and an SLA handbook containing SLA research findings for cloud environment and they are worth reading~\cite{Chana2014}\cite{Mirobi2015}\cite{baset2012}. Moreover, some research focuses on \textit{decision-making} process in SLA for cloud computing such as \cite{Franke2016}\cite{andrzejak2010}. The paradigm of  cloud computing guarantees trustworthy services, which is available through any place within the world, in associate degree on-demand manner. To adopt cloud services, insufficient security has been already known as a significant obstacle. To accommodate the risks related to outsourcing knowledge and applications to the cloud, new ways for security assurance are much required. Therefore, some research has been done on SLA security in the cloud. For instance they propose authentication interface to access a cloud service \cite{Bernsmed2011}\cite{Bajpai2012a}\cite{Bajpai2012b}\cite{buyya201111}. In some published research reports, specific theories have been utilised for SLA design and development. In \cite{Tang2014}, the authors utilised \textit{Bayesian Model} to propose an approach for predicting SLA violations, which uses measured datasets (QoS of used services) as input for a prediction model. Another research used the advantages of the \textit{Game Theory} to model an SLA negotiation and framework for the QoS assurance purpose within the clouds federation~\cite{figueroa2008}\cite{AlFalasi2016}. Furthermore, in~\cite{Shenoy2016} data-driven probabilistic has been considered for modelling application resource demand for resource allocation and they claimed that more than 50\% savings are demonstrated using the proposed approach for resource allocation in the Yahoo's data centres.

Renewable energy plays a crucial role for assistance of meeting basic energy needs through the use of modern technologies known as \textit{GreenTech}. Therefore, with the large-scale deployment of either virtualised or physical data centres, energy consumption and SLA have become the immediate issue to be solved. That is why some researchers aimed to find energy-efficient SLA solutions. However, yet there are small number of published articles in this area and this research domain is trending in recent years \cite{son2017}\cite{GAO2014}\cite{Hassan2017}\cite{Amokrane2015} \cite{Goudarzi2012}\cite{cao2012}. Generally, in terms of SLA architecture there is not sufficient resources and this constraint goes to the SLA architecture for cloud computing as well. In \cite{simao2013}, the authors outlined a value model followed by an SLA with a partial utility-driven scheduling architecture, that comes with the partial utility the consumer offers to a specific level of depreciation. 

SLA-aware resource allocation in cloud environment has gained more interest to pursue research on its open challenges and many research articles have been published in this domain. In \cite{iccke2013}, as an example, a mutual negotiation protocol that is using the alternate offers model has been proposed for resource allocation and scheduling in Grid Federation and relevant issues and challenges of SLA-aware resource allocation are discussed in \cite{Chana2014}. In \cite{GAO2014}, the authors proposed a dynamic resource management based on queuing theory through integer programming, control theory techniques and integrating the timing-analysis to schedule Virtual Machines (VMs) which is driven by range-based non-linear reductions of utility, different for classes of clients and among different ranges of resource allocations \cite{simao2016} and another approach computes the resource allocation ratio based on the historical monitoring data from the online analysis of the host and network utilisation but any pre-knowledge of workloads \cite{son2017}. Furthermore, there are existing works published with virtualised and proactive approaches such as \cite{Watson2010}\cite{Morshedlou2014} and also research challenges and published solutions so far are discussed in \cite{buyya201111}. An SLA based cloud computing that facilitates resource allocation which takes into account geographical location and the workload of distributed data centres is proposed in \cite{Son2013}. An efficient heuristic algorithm based on dynamic programming and convex optimization is introduced to solve the mentioned resource allocation problem by H. Goudarzi \textit{et al.}~\cite{Goudarzi2012} and yet there are several articles published in this domain for cloud computing such as \cite{Singh2016}\cite{Van2009}\cite{Goudarzi2011}\cite{GARG2014}\cite{Thangaraj2010}\cite{Zhao2015}, for further reading.

\subsection{SLA Architecture}\label{sec5.2}

Typical SLAs are only characterised at a single layer and do not provide insight into parameters or metrics at the various service stack lower-layers. Therefore, they do not permit providers to manage their service stack optimally. In \cite{Happe2011}, a reference architecture is proposed for a multilevel SLA management framework. All fundamental concepts of the proposed framework have been discussed in this study and its main architectural interactions and components explained in detail.

Many challenges of SLA-oriented resource allocation have been addressed in data centres to  satisfy  competing  applications requests for computing services. In this study, the authors proposed an SLA-oriented architecture to overcome resource allocation challenges in data centres. They claimed that the proposed framework can be effectively implemented using the proposed ''Aneka platform".

It requires to be emphasised that the SLA architecture context has, so far, not been investigated well and there are not enough articles in this topic. Therefore, we expect that, it would be one of potential future research directions in this context.

\subsection{SLA Monitoring}

One of the SLA objectives is SLA monitoring. An SLA monitoring must be accomplished right after the contractual agreement and SLA deployment to meet the EUs expectations. The service monitoring is significantly important to follow the progress of its performance and to ensure that the service complies with the agreed SLA. The SLA monitoring is accomplished by applying several statistics like monitoring data, analysing, a systematic process of collecting and other factors that derive the higher value from the business. Once done with the monitoring, it proceeds with the reporting of SLA where the SP is capable to clearly see the dashboard breakdowns with time, policy and status where it can recognise the areas of the problem. Both SLA monitoring and reporting always assist to meet the agreement for business applications and provide the highest possible performance.

In~\cite{Chhetri2016}, it is shown that the dynamic, diverse and unforeseeable nature of both application workloads and cloud services make quality-assured provision of such cloud service-based applications (CSBAs) a main challenge. The authors propose a cross-layer framework for SLA monitoring and its main aspects contain:  (a) realtime, fine-grained visibility into CSBA performance,  (b) visual descriptive analytics to identify correlations and interdependencies between cross-layer performance metrics,  (c) temporal profiling of CSBA performance,  (d) proactive monitoring, detection and root-cause analysis of SLA violation, and  (e) support for both reactive and proactive adaptation in support of quality-assured CSBA provision. The proposed approach for SLA monitoring is validated by a prototype implementation.

Joel Sommers \textit{et al.}, in \cite{Sommers2010}, active measurements for the SLA performance metrics are unified in a discrete time-based tool called SLAM (SLA Monitor). The SLAM tool implemented to carry out multi-objective probing with two different topologies. SLAM transmits UDP packets in a one-way manner between a transmitter and receiver. SLA monitoring is of significant interest to both EUs and SPs to ensure that the network is operating within the acceptable bounds. As discussed in the article, the obtained results illustrate that standard techniques for measuring loss rate, delay variation and end-to-end delay cannot provide an accurate approximation of the state of the network, thereby preventing an accurate assessment of SLA compliance. Among event based SLA monitoring,~\cite{leitner2010} proposed a monitoring approach called prevention and prediction based on an event monitoring (PREVENT) system, a framework for prediction of run-time and further prevention of violations. PREVENT is based on the idea of monitoring as well as analysing run-time data to trigger adaptation actions in endangered composition instances.

The ''Web Service Level Agreement (WSLA)" framework is targeted at describing and monitoring SLAs for Web services. Although, WSLA has been developed for an environment of Web services, it is also enforceable to any other inter-domain management scenario, like service management and business process, or the management of  systems, applications and networks in general. Upon the receipt of an specification of SLA, automatically the WSLA monitoring services are configured to enforce the SLA.  An implementation of the WSLA framework termed SLA Compliance Monitor, is publicly available as part of the IBM Web Services Tool-kit \cite{Keller2003}.

The SLA monitoring is playing a key role to ensure compliance with the stated terms of the contract.  In order to make an appropriate system for service levels measurement, it is very important to realise what is realistically feasible as far as service is concerned. The technical metrics do have a role to play in consideration of service levels. The needs to be considered in this stage soonest response time, scalability, delay and jitter for QoS and other aspects required to be taken into account for QoX in future.


\section{Open Problems and Future Research Direction}\label{sec6}

To make reading the SLA-related open problems easy, we proposed five categories to cover existing challenges: a) Negotiation procedures, b) Methodology and modelling, c) Implementation and deployment, d) SLA monitoring and e) Security. 

\subsection{The SLA Negotiation Procedures}

Five specific open challenges are identified for the SLA negotiation protocols and processes. First and foremost, SLA negotiation processes and protocols  diversity constrain the negotiation for establishing new SLAs as it was initially proposed by \cite{linlin2010}\cite{Wu2010}\cite{Yan2007}. Moreover, as the SLA is the pre-set agreement among the actors, the implemented SLA modification, and SLA negotiation between separated administrative domains are not an easy process and most likely it is impossible~\cite{linlin2010}\cite{Wu2010}. 

Often SLAs are technical documents regarding to terminology and concepts which may only be understood by a minor class of technology oriented specialists. Therefore evaluation and improvement do not take place on a regular basis. Such "dead-end'' documents have a very restricted meaning for EUs and their management as it was initially proposed by \cite{Trienekens2004}. On the other hand, unclear service specifications can be considered as an open challenge and yet there is a need of protocol to develop a comprehensive SLA. For instance, agreements on "the availability of a network'' are generally determined using a metric called the percentage of availability. It is extremely tough to specify what the accurate meaning of such a metric is in the context of a specific business location~\cite{Trienekens2004}. Although some recent articles proposed frameworks considering SLA with accuracy such as \cite{Rahim2015}, yet further research is required to be done. By providing such accuracy in the agreement, could the provider answer what is the difference and/or differences between an availability percentage of 98\% and 99\%? The agreement can also define well whether this 98\% is on a yearly basis or not. If it is not clearly defined in the SLA, it means that the network is allowed to go "Down'' for a whole week, after being "Up'' for the last 51 weeks!

\subsection{Methodology \& Modelling}

SLA is a bilateral agreement between two actors and each involved party must have a level of satisfaction achieved in this negotiation (the win-win scenario). The admission control policy is another open problem in SLA methodology and modelling development that needs more investigation and research because, from the SP point of view, a decision on which client demand to accept affects the reputation of the SP,  profit and performance \cite{linlin2010}\cite{Wu2010}. 

Most of proposed methods developed are based on either arbitrary assumptions and parameters or very general parameters such as 99.99\% availability percentage! These would be fine for overall evaluation and/or documentation purpose, but the problem reveals when it goes for a real-time implementation. Therefore, definition of more realistic SLAs should be researched because it may lead to a better use of the network assets and resources. 

With the variety of services provided by SPs, besides the QoS, other quality requirements, named QoX, such as Quality of Experience (QoE), Quality of Information (QoI), discussed in Section ~\ref{sec3}, must be taken into the account in new SLA modelling and establishment \cite{german2016}.

\subsection{Implementation and Deployment}

There are both practical and theoretical constraints to SD-SLA (Software-Defined SLA) \cite{german2016}\cite{arkham2017}\cite{STANIK2014}\cite{son2017}. Beyond the cost and physical limitations that are always system-level parameters that need to be managed, poorly designed SD-SLAs may not be able to be implemented on software-defined networks to support provided services such as cloud services. As an instance, if essential operations are serialised, afterwards, they cannot be programmatically scaled out and up to satisfy an SD-SLA. By developing and implementing appropriate SD-SLAs, there are chances to step into a continuous model for many significant background-processes, that are previously required to be scheduled due to the limitations of fixed resources \cite{Lango2014}. Furthermore, yet there are further opportunities to develop a programmatic SD-SLA validation through automated test analytic and infrastructure \cite{Lango2014}\cite{Bouchenak2013}.

The performance forecast management enables the recommendation for performance improvement and optimisation, therefore it can be considered as an open question in utility computing environments. Moreover, dynamic management of resource allocation has to be considered in the implementation of SLAs, because it addresses which resource is the best and appropriate for a current admitted request from the point of view of both actors \cite{linlin2010}\cite{Wu2010}.

\subsection{SLA Monitoring}

SLA monitoring measures an SLA compliance and monitors true uptime for provided services which is essential for better service delivery and network monitoring is utilised to ensure the hosts and nodes connections with the specified bandwidth as well as monitoring the proper packet delivery. The SLA should be established between providers and end-users from a different end-to-end point of view. As an instance, if the service of system has been outsourced, not only from SP to a EU but also from one SP to another SP, there must be an SLA agreement between them and this requires a dynamic SLA monitoring solution as well to help in managing the network and to maintain the SLA requirements~\cite{Marilly2002}\cite{linlin2010}\cite{Sommers2007}.

\subsection{Security}

Security capabilities measurement which can be guaranteed and quantified for SLA contexts is one of important challenges in expressing security properties. Capability of security is a combination of mutually-reinforcing security controls to mitigate risks as well as demonstrate compliance with the EUs security requirements
\cite{nugraha2017}\cite{nugraha20171}\cite{ross2014}. 

The lack of trust established in SLA agreement about particular security capabilities offered and guaranteed by the SPs is an important challenge faced by SLA actors. Several certifications have been proposed in cloud-computing and outsourcing environments. Nonetheless, these certification schemes are largely inappropriate in the service provisioning context and they do not ensure better security \cite{nugraha2017}\cite{nugraha20171}\cite{Anderson2008}.

In the context of SLAs, metrics can be defined and utilised to measure and track the compliance of security-related SLAs. However, security-related SLAs' metrics are not well-established so far, and available SLA metrics are typically measured and defined based on implemented and statistics  with appropriate provisions and capabilities related to EUs and SPs. Therefore, effective security metrics development for SLAs has proven to be very challenging \cite{nugraha2017}\cite{Schneier2000}\cite{Takahashi2013}.

Table~\ref{openproblems} highlights and classifies the identified open problems in SLA research domain and five categories are addressed in the table as SLA Negotiation, Methodology \& Modelling, Implementation \& Deployment, SLA Management \& Monitoring and Security.

\begin{table}[!htbp]
\begin{center}
\caption{SLA open problems}
\label{openproblems}
\footnotesize
\begin{tabular}{p{.5cm}|p{4cm}|p{1.7cm}}

\hline \hline 
Cat. & Open Problems & References\\
\hline \hline

 \multirow{8}*{\rotatebox[origin=c]{90}{SLA Negotiation}}& - The negotiation constrains for establishing SLA & \cite{linlin2010}\cite{Wu2010}\cite{Yan2007} \\
 & - An implemented SLA modification & \cite{linlin2010}\cite{Wu2010}\\
    & - SLA negotiation among different administrative domains & \cite{linlin2010}\cite{Wu2010} \\
    & - ''Dead-end'' SLA documents & \cite{Trienekens2004} \\
    & - Unclear service specifications & \cite{Trienekens2004} \\
\hline

\multirow{5}*{\rotatebox[origin=c]{90}{\specialcell{Methodology \\ Modelling}}}& - Lack of other quality requirements (QoX) consideration & \cite{german2016}\\
& - Lack of a win-win deal commitment for all network actors  & \\
& - Admission  control policies & \cite{linlin2010}\cite{Wu2010} \\
& - Definition of more realistic SLAs  & \cite{linlin2010}\cite{Wu2010} \\
\hline

\multirow{7}*{\rotatebox[origin=c]{90}{\specialcell{Implementation\\ Deployment}}} & - Implementation in software-defined network architecture & \cite{german2016}\cite{arkham2017} \cite{STANIK2014}\cite{son2017}\cite{Lango2014}\\
 & - Performance forecast management in utility computing environments & \cite{Lango2014}\cite{Bouchenak2013}\\
 & - The efficient resource allocation management  & \cite{linlin2010}\cite{Wu2010}\\
\hline

\multirow{6}{*}{\rotatebox[origin=c]{90}{\specialcell{Management\\ Monitoring}}} & - SLA has to be established among the providers and clients from various end-to-end point of view. & \cite{Marilly2002}\cite{linlin2010}\cite{Wu2010}\\
&&\\
&&\\
\hline

 \multirow{6}{*}{\rotatebox[origin=c]{90}{Security}} & - Quantifying security properties in SLA contexts & \cite{nugraha2017}\cite{nugraha20171}\cite{ross2014}\\
& - Specifying capabilities of security in SLA contexts &  \cite{nugraha2017}\cite{Anderson2008}\\ 
 & - Evaluating security capabilities specified in SLA contexts & \cite{nugraha2017}\cite{Schneier2000}\cite{Takahashi2013}\\
\hline

    \end{tabular}
  \end{center}
\end{table}


\section{Conclusions}\label{sec7}
The increasing use of ICT services makes the establishment of efficient SLA vital for InP, SP and EU to maintain ongoing contracts. In this survey a large number of research studies, models and methods were described with the aim of revealing the importance of research activities in the SLA domain and it shows that there is a large research effort to propose fair SLA methods. Based on the reviewed articles the current SLA research state of art has been identified in the domain and a comprehensive SLA conceptual-map was developed to identify SLAs related concepts. 
The SLA elements and actors were introduced and the importance of their key roles on achieving a successful SLA have been discussed followed by SLA characteristics and SLO. This survey was organised in a way to provide a better understanding of existing methods, models and architectures for readers with different research interests related to SLA and a reading-map is provided in Section~\ref{sec1}. Furthermore, SLA relevant application domains are identified and discussed to show their applicable scopes in ICT industries.

Although the proposed methods and architectures claimed that they were successful in satisfying an actor with a focus on particular service, yet the lack of a generic SLA method requires further research to achieve win-win SLA deployment among all involved actors. This survey pointed out existing research gaps in utilising SLA elements to develop a generic methodology for SLAs establishment, deployment, management and particularly that will lead to a win-win situation for all involved actors.

Finally, various open research problems and future research directions were discussed and classified in five categories in order to encourage researchers in further research on developing generic SLA models and architectures. This study is part of an ongoing SLA project and is an active research on how to specify and develop SLAs to achieve win-win agreements among all actors.



\appendix

\section{SLA Conceptual-Map}\label{app1}

A concept map or conceptual-map could be a diagram that depicts prompt relationships between ideas. It is a graphical tool that is utilised to organise and to structure knowledge. A conceptual map usually represents concepts and information. The connection between ideas are often articulated by means of predicates. The conceptual maps are developed to guide specific researches. Therefore, for a same topic, SLA for example, different conceptual maps may be developed to achieve different research objectives.

\begin{figure}[!htbp]
\centering
\includegraphics[draft=false,width=7.1in,center]{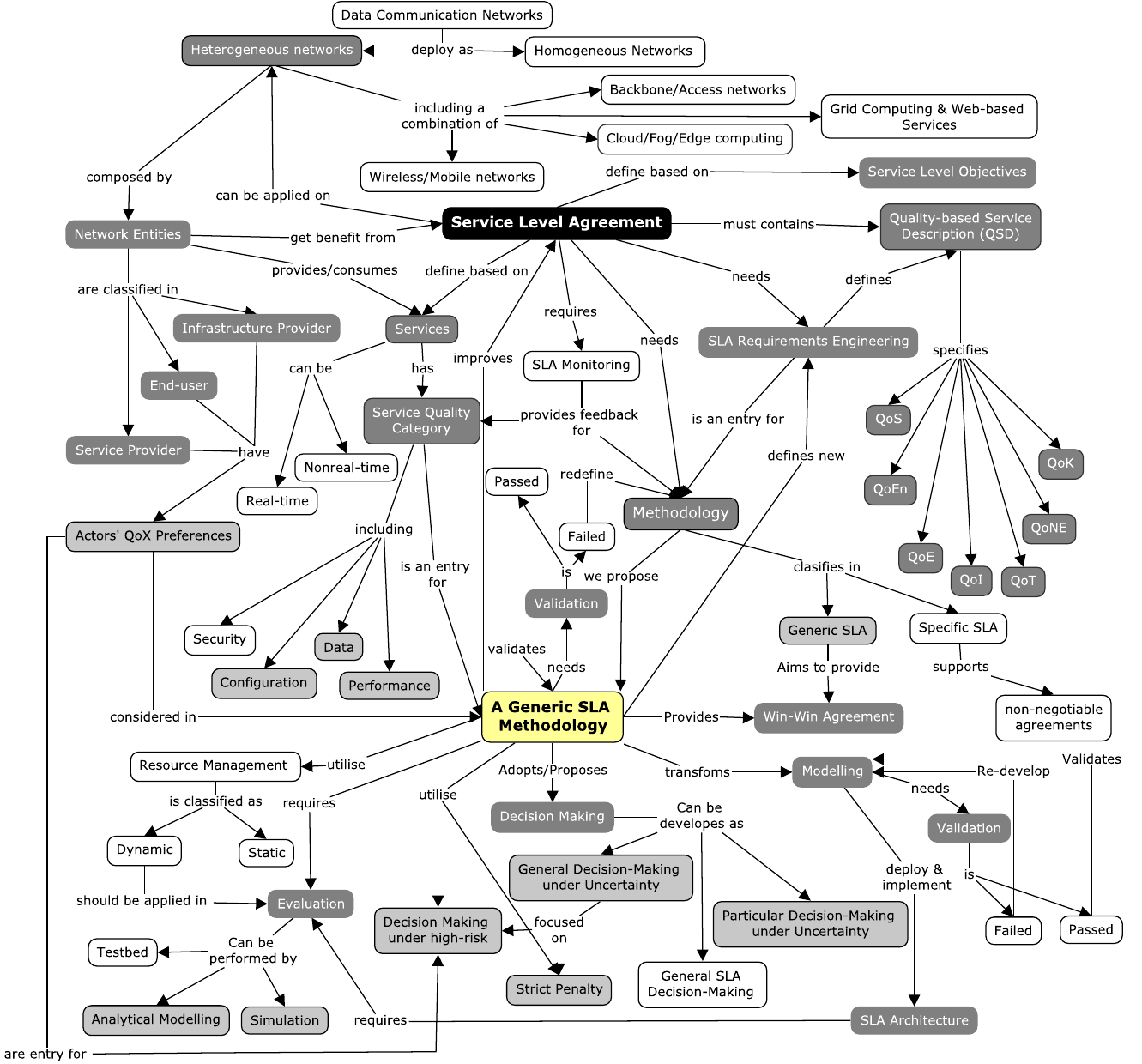}
\caption{SLA conceptual-map created using CMap~\cite{cmap} }
\label{SLACmap}
\end{figure}

Figure~\ref{SLACmap} illustrates the proposed conceptual-map to achieve a comprehensive and a generic SLA methodology for computer and communication networks, and to take all quality requirements into account in SLA deployment. The map starts in the top centre box identified as \textit{Data Communication Networks} and it is composed by network entities that can get benefits of SLA to provide  promised level of service for EU and/or SP. To achieve a desirable SLA, the model must be defined based on SLOs and contains QSDs. In this proposed conceptual-map, all relevant concepts such as SLA monitoring and requirements engineering are considered to provide necessary feedbacks and to supply required information as entries for an SLA methodology development.

In the proposed map, all relevant concepts are organised in a way to achieve \textit{A generic SLA methodology} (shown in yellow box in the middle of the map) that can provide a win-win agreement among all actors by considering their QoX requirements. A generic SLA methodology requires suitable    resource management and decision-making algorithms. Based on a generic SLA methodology, an SLA model and architecture can be proposed. Eventually, a model and architecture of SLA can be evaluated and validated before getting deployed. 

\newpage
\onecolumn
\newpage
\twocolumn

\bibliography{Survey}
\end{document}